\documentclass[twocolumn,tighten,times]{aastex631}

\usepackage{amsmath}
\usepackage{graphicx}
\usepackage{hyperref}
\graphicspath{{./}}

\newcommand{\mgii}{Mg\,{\sc ii}}
\newcommand{\nev}{[Ne\,{\sc v}]}
\newcommand{\heii}{He\,{\sc ii}}
\newcommand{\oiii}{[O\,{\sc iii}]}
\newcommand{\nii}{[N\,{\sc ii}]}
\newcommand{\sii}{[S\,{\sc ii}]}
\newcommand{\ha}{H$\alpha$}
\newcommand{\hb}{H$\beta$}
\newcommand{\msun}{M_\odot}
\newcommand{\kms}{km\,s$^{-1}$}

\shorttitle{DESI MgII Absorption and Stellar Surface Density}
\shortauthors{Rong \& Liu}

\begin{document}

\title{Stellar Surface Density Modulates \mgii\ Cool-gas Outflow Absorption in
DESI Star-forming Galaxies}

\correspondingauthor{Yu Rong}

\author{Yu Rong}
\email{rongyua@ustc.edu.cn}
\affiliation{Department of Astronomy, University of Science and Technology of China, Hefei, Anhui 230026, China}
\affiliation{School of Astronomy and Space Sciences, University of Science and Technology of China, Hefei 230026, Anhui, China}

\author{Shihong Liu}
\affiliation{Department of Astronomy, University of Science and Technology of China, Hefei, Anhui 230026, China}
\affiliation{School of Astronomy and Space Sciences, University of Science and Technology of China, Hefei 230026, Anhui, China}

\begin{abstract}
Galaxy outflows are usually ordered by stellar mass and star-formation rate
(SFR), but the same feedback budget may couple differently to gas in diffuse
and compact galaxies.  We use Dark Energy Spectroscopic Instrument (DESI)
Data Release 1 stacked spectra of massive star-forming galaxies at
$0.35<z<1.0$ to test whether stellar surface density,
\(\Sigma_\star=M_\star/(2\pi R_e^2)\), is an independent empirical coordinate
of down-the-barrel singly ionized magnesium (\mgii) cool-gas absorption.  In
AGN-clean samples matched in stellar mass, and in a stricter sample matched in
both stellar mass and a Balmer-line SFR proxy, the \mgii\ outflow equivalent
width (EW) rises monotonically with \(\Sigma_\star\) in every redshift bin.
From the lowest to highest \(\Sigma_\star\) tertile, EW$_{\rm out}$ increases
by \(0.37\)--\(0.61\)~\AA, while the absolute outflow velocity changes only
weakly.  DESI therefore shows that cool-gas outflow strength in massive
star-forming galaxies is not set only by how much stellar mass or star
formation a galaxy has, but also by how tightly the galaxy is built.  The
structural dependence points to changes in the absorbing velocity distribution
and/or the effective covering fraction of cool outflowing gas.
\end{abstract}


\section{Introduction}

Galactic winds are a major channel through which stellar feedback regulates
galaxy growth, transports metals, and connects the interstellar medium to the
circumgalactic medium
\citep[e.g.,][]{chevalier1985,dekel1986,heckman2000,veilleux2005,rupke2005,
murray2005,oppenheimer2006,muratov2015,somervilledave2015,naab2017,
veilleux2020}.  Observationally, outflows are seen from the nearby Universe to
the peak epoch of cosmic star formation through optical emission-line
kinematics, Na\,{\sc i} absorption, molecular and atomic gas tracers, and
rest-frame ultraviolet interstellar absorption
\citep{heckman2000,shapley2003,martin2005,rupke2005,steidel2010,
arribas2014,cicone2014}.  These measurements provide the empirical basis for
the feedback prescriptions used in galaxy-evolution models, but they also show
that winds are multi-phase, geometrically complex, and difficult to summarize
with a single velocity or mass-loading factor.

The cool, low-ionization phase is important because it contains a large
fraction of the entrained neutral gas and metals.  In integrated spectra,
the down-the-barrel geometry gives a direct view of this material projected
against the galaxy's own stellar continuum.  The \mgii\
$\lambda\lambda2796,2803$ and Fe\,{\sc ii} transitions are standard tracers of
cool outflowing gas at intermediate redshift: they are strong, accessible from
the ground over a broad redshift interval, and sensitive to gas at velocities
of a few hundred \kms.  Large surveys and deep
spectroscopic samples have shown that blueshifted low-ionization absorption is
common in star-forming galaxies, and that its incidence, equivalent width, and
velocity depend on stellar mass, SFR, luminosity, inclination, and starburst
properties
\citep{tremonti2007,weiner2009,rubin2010,martin2012,kornei2012,
erb2012,bordoloi2014,rubin2014,chisholm2015,heckmanborthakur2016}.  At the
same time,
\mgii\ profiles are not simple linear measures of outflow column density:
saturation, partial covering, resonant emission filling, and velocity
structure all shape the observed absorption profile
\citep{rubin2011,erb2012,martin2013,rubin2014}.

Most empirical scaling relations have been framed in terms of integrated
quantities such as stellar mass, $M_\star$, and star-formation rate, SFR.
However, feedback should also depend on how concentrated the stars and star
formation are.  Compact starbursts can drive unusually fast outflows,
suggesting that feedback depends on the spatial arrangement of stars and gas as
well as on the integrated star-formation budget
\citep{diamondstanic2012,sell2014,heckmanborthakur2016}.  Local starburst
work, integral-field-unit (IFU) studies, and high-redshift emission-line
measurements have likewise emphasized that large-scale winds are connected to
how intensely feedback is concentrated within galaxies
\citep{heckman2000,newman2012,arribas2014}. 
These results suggest that total SFR alone is an incomplete feedback
coordinate.  A galaxy with the same $M_\star$ and SFR but a smaller effective
radius may couple energy and momentum to cool gas differently, either because
the launch region is more compact or because the stellar potential and gas
column structure are different.

The relevant structural variable for a large statistical experiment is a
surface-density proxy.  We use
$\Sigma_\star=M_\star/(2\pi R_e^2)$, where $R_e$ is the effective radius,
because it is available for a very large DESI sample and captures the
compactness of the stellar body.  This choice is complementary to SFR-based
wind scalings: $M_\star$ sets much of the potential-well scale, SFR traces the
instantaneous feedback budget, and $\Sigma_\star$ asks whether the same mass
and feedback are arranged in a way that changes the observed cool outflow.
Testing this requires both large numbers and controlled comparisons, since
$\Sigma_\star$, $M_\star$, and SFR are mutually correlated in real galaxy
samples \citep{Franx2008,Wuyts2011,Rong20}.

The Dark Energy Spectroscopic Instrument (DESI) now provides the sample size
needed to make such a differential test.  In previous work, stacked DESI
spectra have already shown that \mgii\ outflow equivalent width (EW) is closely
linked to SFR while the characteristic velocity is more closely tied to stellar
mass \citep{yu2025}.  That result motivates the next structural question: at
fixed $M_\star$ and SFR, does a more compact galaxy show stronger cool
outflowing gas?  We address this question with a controlled DESI stacking
experiment: in each redshift bin we compare \mgii\ absorption across
$\Sigma_\star$ tertiles after matching the stellar-mass distribution and, in a
stricter test, the line-SFR distribution.  The resulting measurements provide
a benchmark for future observational and theoretical work that asks whether
galaxy structure, beyond \(M_\star\) and SFR, regulates cool-gas feedback.

\section{Data and Sample}

\subsection{Catalogs and Basic Cuts}

We use DESI Data Release 1 (DR1) coadded spectra
\citep{desispectrograph,desidr1} and the DESI stellar-mass and
emission-line value-added catalog of \cite{Zou24}.  The catalog uses the official DESI
pipeline redshifts; the catalog measurements used in this work are stellar-population
quantities, imaging sizes, and emission-line fluxes.  We require main-survey
galaxies with redshift $0.35<z<1.0$,
$\log(M_\star/\msun)>10.5$, finite positive Legacy Survey size parameter
$R_e$, a valid coadd file, DESI \texttt{ZWARN}$=0$, and DESI spectral type
`GALAXY', and $|c(z_{\rm catalog}-z_{\rm DESI})/(1+z_{\rm DESI})|<100$~\kms, where $c$ is the speed of light, $z_{\rm catalog}$ is the value propagated in the catalog of \cite{Zou24}, and $z_{\rm DESI}$ is the official DESI pipeline redshift. The last cut removes sources with obvious redshift failures before rest-frame stacking.

The stellar surface density is defined as
\begin{equation}
 \Sigma_\star = \frac{M_\star}{2\pi R_e^2},
\end{equation}
where $R_e$ is the physical half-light radius inferred from the catalog
\texttt{SHAPE\_R} measured by the Legacy Surveys imaging model fits
\citep{dey2019} and converted to kpc using the official DESI redshift.
\texttt{SHAPE\_R} is the observed half-light/effective radius of the galaxy
model in arcsec.  

\subsection{AGN-clean Selection}

We remove possible active galactic nucleus (AGN), composite,
low-ionization nuclear emission-line region (LINER)-like, and
shock-contaminated systems before stacking.  Where \ha, \hb,
\oiii$\lambda5007$, and \nii$\lambda6583$ have S/N$>3$, we use the
\nii\ Baldwin--Phillips--Terlevich (BPT) diagram and veto sources outside the
star-forming sequence \citep{baldwin1981,kauffmann2003,kewley2001}.
Where \sii$\lambda\lambda6716,6731$ is also measurable, we apply the analogous
\sii\ BPT veto following the standard optical classification framework
\citep{kewley2006}.  At redshifts where the full BPT set is not always
available, we additionally remove sources with \nev$\lambda3346$,
\nev$\lambda3426$, or \heii$\lambda4686$ detected at S/N$>2$.  Any available
indication of non-stellar hard ionization removes the object.  This selection
is stricter than a completeness-oriented sample, but it yields cleaner
star-forming stacks for testing stellar-feedback-driven absorption.

\subsection{Redshift and Surface-density Bins}

We analyze three redshift bins:
$0.35<z<0.45$, $0.45<z<0.60$, and $0.60<z<1.00$.  In each redshift bin, the AGN-clean
sample is split into tertiles of $\log\Sigma_\star$.  We construct two
controlled samples.  In the first, the three $\Sigma_\star$ tertiles are
randomly down-sampled in narrow $M_\star$ bins to the same mass distribution.
This mass-matched sample preserves the observed covariance between
$\Sigma_\star$ and SFR.  In the second, the tertiles are matched in both
$M_\star$ and line-SFR using two-dimensional $M_\star$--SFR cells. The line-based SFR is estimated from \hb\ when its signal-to-noise ratio,
S/N(\hb), is greater than 3, otherwise from \ha\ when S/N(\ha)$>3$ and the line
falls in the DESI wavelength range \citep{kennicutt2012}.  This
sample tests whether a residual $\Sigma_\star$ signal remains after SFR is
also fixed.  

\begin{figure*}
\centering
\includegraphics[width=0.98\textwidth]{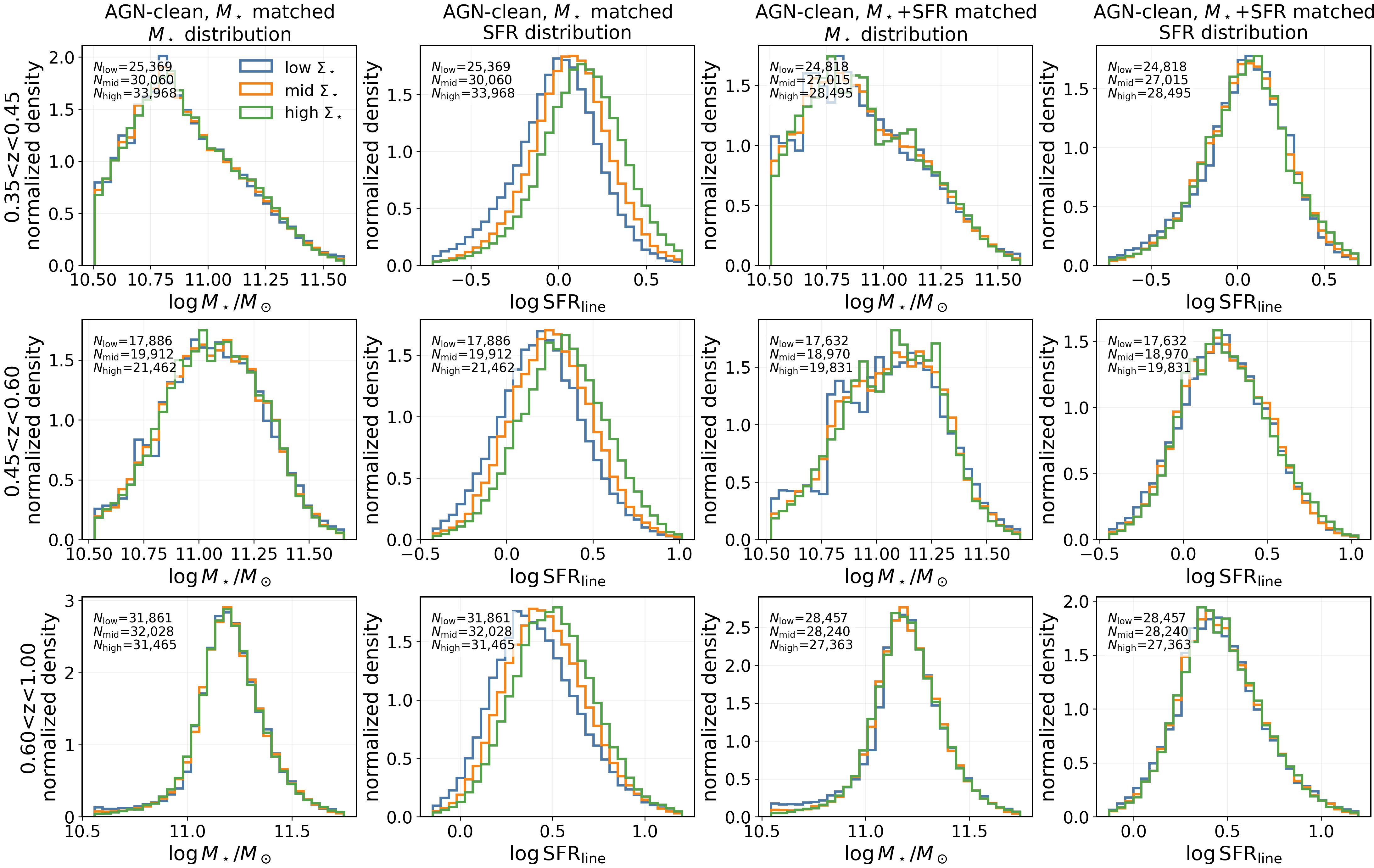}
\caption{Control distributions for the two final AGN-clean samples.  Rows show
the three redshift intervals, and colors show the low-, middle-, and
high-$\Sigma_\star$ tertiles; the annotated numbers give the final number of
galaxies entering each stack.  The left two columns show the sample matched in
$M_\star$ only, preserving the observed covariance between $\Sigma_\star$ and
SFR.  The right two columns show the sample matched in both $M_\star$ and
line-SFR.}
\label{fig:controls}
\end{figure*}

The per-tertile sample sizes are 51,012, 23,982, and 34,207 for the
mass-matched sample, and 46,063, 23,071, and 30,164 for the
mass+SFR-matched sample in the three redshift bins.  Figure~\ref{fig:controls}
shows that the matching works as intended.  In the mass-matched sample the
median $M_\star$ range across $\Sigma_\star$ tertiles is only
$0.0002$--$0.003$ dex, while SFR still shifts systematically with
$\Sigma_\star$.  In the mass+SFR-matched sample the median SFR range is also
reduced to $0.001$--$0.006$ dex.

\section{Stacking and Measurements}
\label{sec:measurements}

Each coadded spectrum is shifted to the rest frame and resampled onto a common grid over 2700--2910~\AA.  We then
continuum-normalize each spectrum using the same local-polynomial approach as
\citet{yu2025}.  Specifically, we mask the strong rest-UV features near
Fe\,{\sc ii}~$\lambda\lambda2586,2600$, \mgii$\lambda\lambda2796,2803$, and
Mg\,{\sc i}~$\lambda2853$, smooth the remaining spectrum with a median filter,
fit a fifth-order polynomial to the unmasked pixels, and divide the spectrum
by this fitted continuum.  Spectra with failed continuum fits or pathological
normalizations are discarded.  Within each bin we median-stack the normalized
spectra with equal galaxy weight; no flux, luminosity, or S/N weighting is
applied.

We define the outflow equivalent width, EW$_{\rm out}$, following
\citet{yu2025}.  We integrate the \mgii$\lambda2796$ blue side over
\(-700<v_{2796}<0\)~\kms\ to obtain EW$_{\rm blue,2796}$, and the
\mgii$\lambda2803$ red side over \(0<v_{2803}<400\)~\kms\ to obtain
EW$_{\rm red,2803}$.  The outflow component is then
\begin{equation}
{\rm EW}_{\rm out}=2({\rm EW}_{\rm blue,2796}-{\rm EW}_{\rm red,2803}).
\end{equation}
EW$_{\rm red,2803}$ is therefore a reference window for subtracting the
non-outflow component, not a doublet-ratio measurement.  We compute
\(v_{\rm out}\) with the same prescription.  Let
\(A(v)=\max[1-F(v),0]\) be the positive absorption depth in the two windows,
with velocities defined relative to the corresponding doublet component.  The
absorption-weighted velocity of the two-window profile, evaluated on the
rest-frame wavelength grid, is
\begin{equation}
v_{\rm tot} =
\frac{
\int_{-700}^{0} v_{2796} A_{2796}(v)\,d\lambda
+\int_{0}^{400} v_{2803} A_{2803}(v)\,d\lambda}
{\int_{-700}^{0} A_{2796}(v)\,d\lambda
+\int_{0}^{400} A_{2803}(v)\,d\lambda}.
\end{equation}
Following \citet{yu2025}, we then define
\begin{equation}
\begin{aligned}
v_{\rm out} &= v_{\rm tot}
\frac{{\rm EW}_{\rm tot}}{{\rm EW}_{\rm out}},\\
{\rm EW}_{\rm tot}
&=2({\rm EW}_{\rm blue,2796}+{\rm EW}_{\rm red,2803}).
\end{aligned}
\end{equation}
The normalization and velocity windows are fixed for all bins.  Uncertainties
on EW$_{\rm out}$ and $v_{\rm out}$ are estimated with
galaxy-level bootstrap resampling.  For each of the 18 stacks
(two controlled samples, three redshift bins, and three $\Sigma_\star$
tertiles), we resample the contributing galaxies with replacement 200 times,
repeat the stack and measurement, and quote the 16th--84th percentile interval.

\begin{figure*}
\centering
\includegraphics[width=0.92\textwidth]{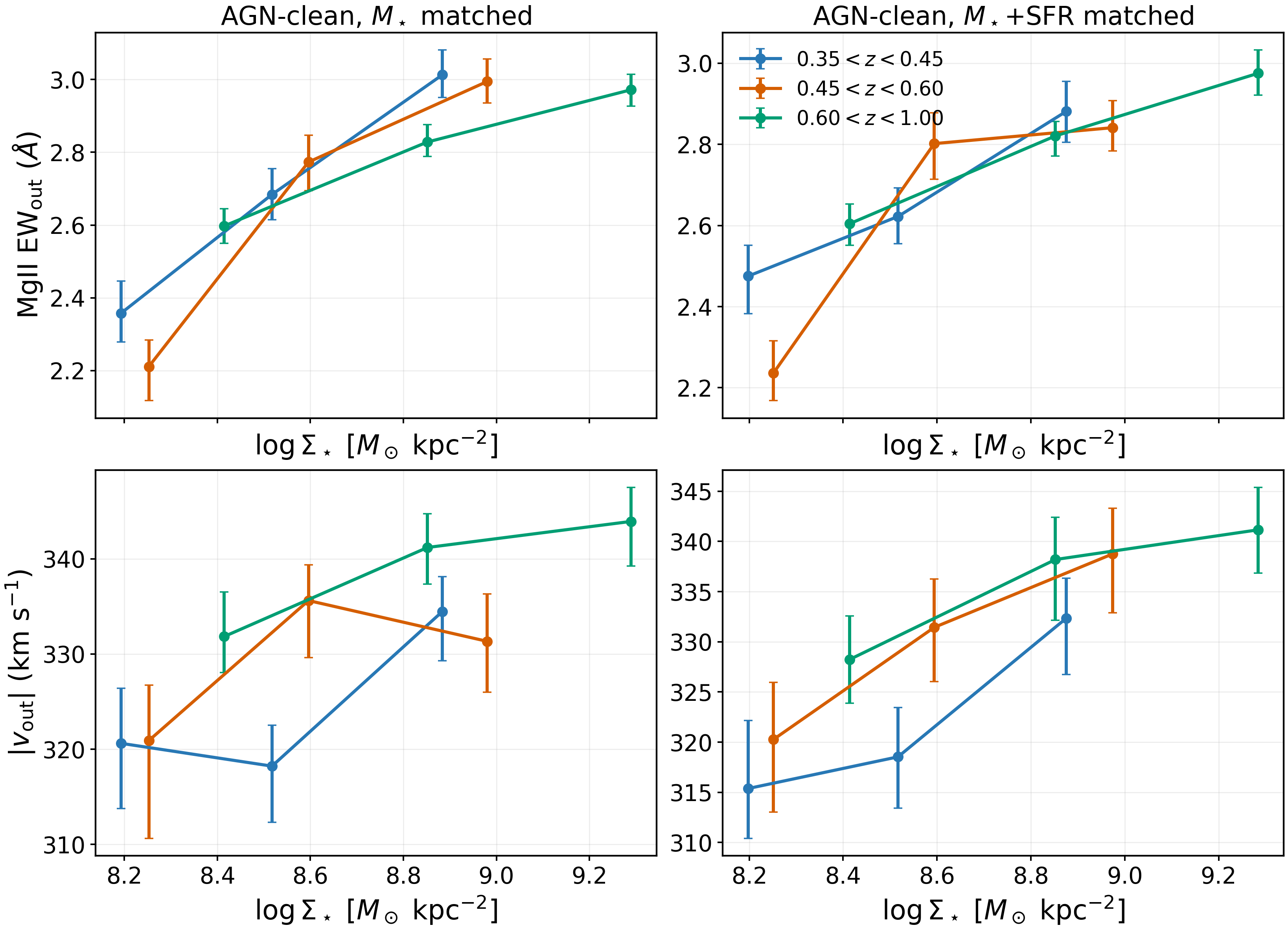}
\caption{Observed \mgii\ outflow absorption strength as a function of stellar
surface density.
Left: AGN-clean stacks matched in $M_\star$.  Right: AGN-clean stacks matched
in both $M_\star$ and SFR.  Points show the three $\Sigma_\star$ tertiles in
each redshift interval.  Error bars are 16th--84th percentile intervals from
200 galaxy-bootstrap realizations per stack.  The EW trend remains after
controlling both $M_\star$ and SFR, while the velocity trend is weaker.}
\label{fig:trends}
\end{figure*}

\section{Results}

\begin{figure*}
\centering
\includegraphics[width=0.92\textwidth]{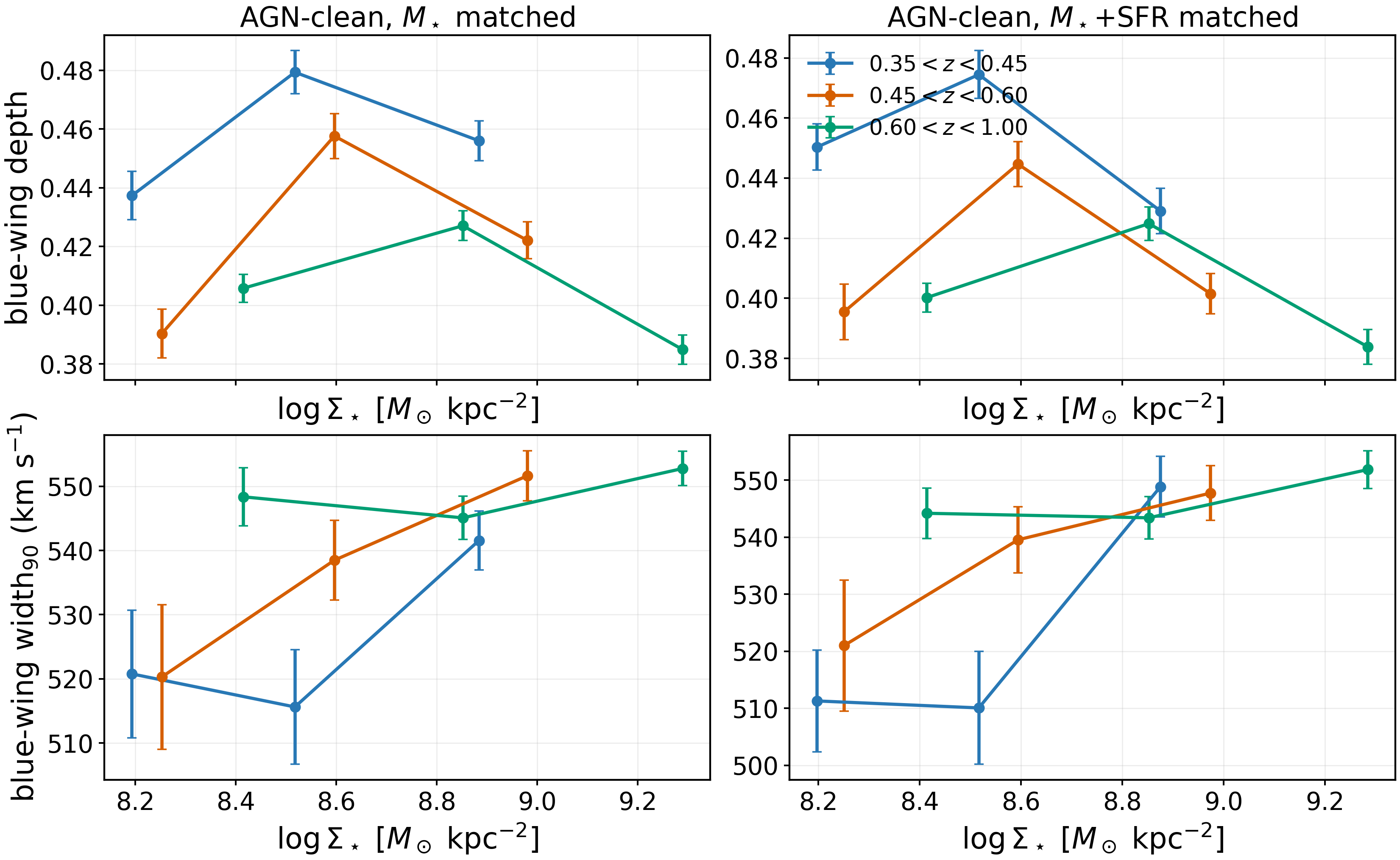}
\caption{Profile-shape diagnostics for the EW$_{\rm out}$ trend shown in
Figure~\ref{fig:trends}.  From top to bottom: blue-wing absorption depth
\(1-P_{10}[F(v)]\) and blue-wing velocity width width$_{90}$.  Error bars are
Monte Carlo uncertainties from the stacked-spectrum error arrays.  The depth
trend is not monotonic, while width$_{90}$ generally increases with
$\Sigma_\star$.}
\label{fig:mechanism}
\end{figure*}

\begin{figure*}
\centering
\includegraphics[width=0.92\textwidth]{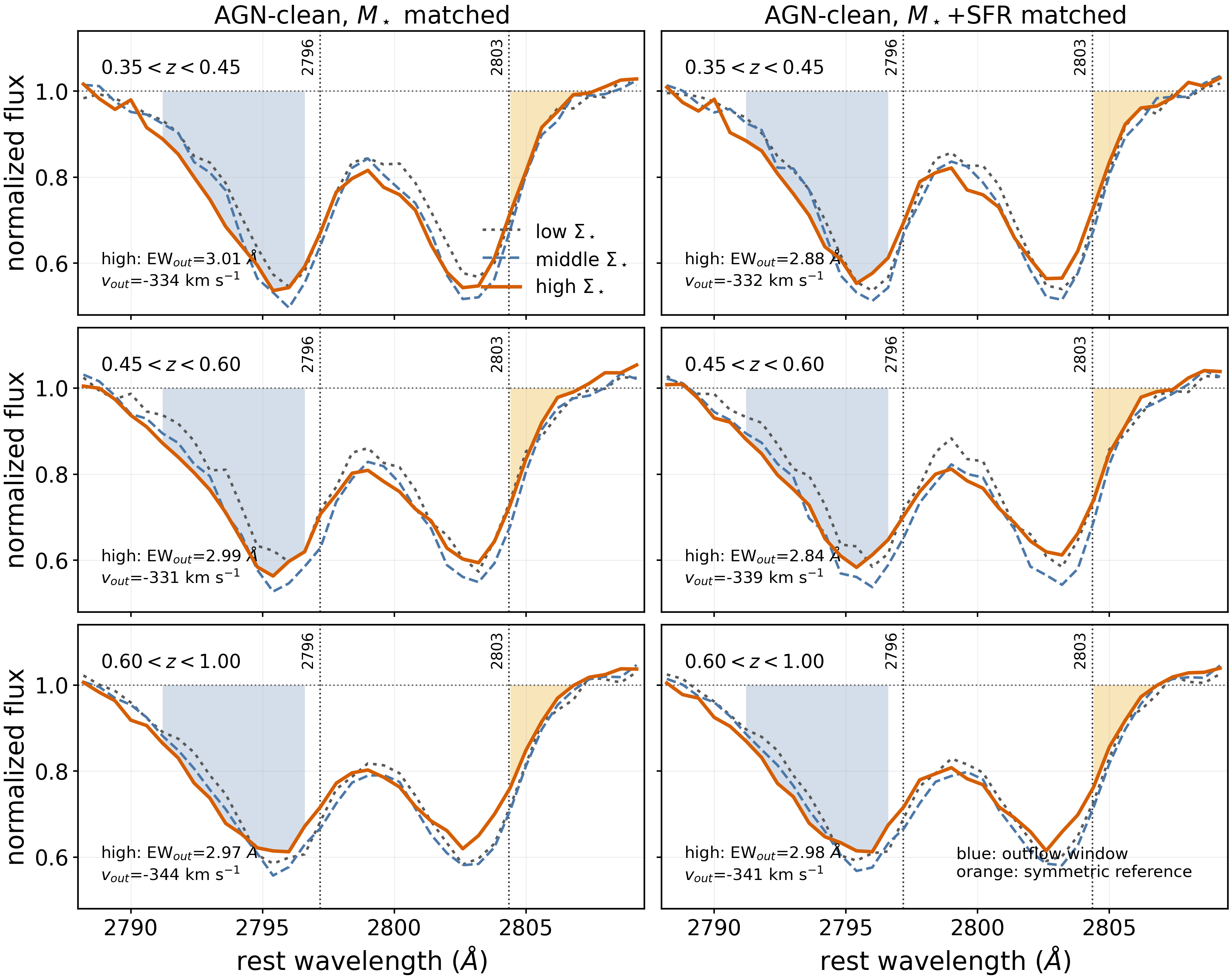}
\caption{Stacked \mgii\ doublet profiles for the two controlled samples.
Line style denotes $\Sigma_\star$ tertile.  The vertical dotted lines mark the
rest wavelengths of \mgii$\lambda2796$ and \mgii$\lambda2803$.  The blue
shaded region marks the blueshifted outflow window for the high-$\Sigma_\star$
stack, while the orange shaded region marks the red-side symmetric reference
window used to estimate the non-outflow component.  The stronger
high-$\Sigma_\star$ absorption is visible directly in the blue wing.}
\label{fig:profiles}
\end{figure*}

\subsection{MgII Outflows Strengthen with Stellar Surface Density}

Figure~\ref{fig:trends} shows the main result.  In both controlled samples and
all three redshift bins, EW$_{\rm out}$ increases from the low- to
high-$\Sigma_\star$ tertile.  In the mass-matched sample, which preserves the
observed SFR--$\Sigma_\star$ covariance, the low-to-high increase is
$0.658^{+0.115}_{-0.104}$, $0.792^{+0.091}_{-0.098}$, and
$0.373^{+0.062}_{-0.060}$~\AA~in the three
redshift bins, respectively.  In the stricter mass+SFR-matched sample,
the increase remains $0.413^{+0.113}_{-0.113}$,
$0.614^{+0.096}_{-0.098}$, and $0.369^{+0.068}_{-0.067}$~\AA\ in the three
redshift bins.  Thus the structural trend remains after controlling both
$M_\star$ and the line-SFR proxy.

The velocity trend is much weaker.  In the fiducial mass+SFR-matched sample,
$|v_{\rm out}|$ increases by
$15^{+8}_{-7}$, $19^{+8}_{-8}$, and $12^{+7}_{-6}$~\kms\ from the low- to
high-$\Sigma_\star$ tertile.  These differences are far smaller than the
single-spectrum DESI instrumental width at the observed \mgii\ wavelengths
(tens to more than 100~\kms, depending on camera and redshift), and should not
be interpreted as a resolved velocity separation between stacks.  The primary
result is the EW increase; the velocity change is very small and may not be trusted.

\subsection{What Drives the EW Trend?}

Figure~\ref{fig:mechanism} shows two non-parametric profile-shape diagnostics
used to interpret the EW trend.  Absorption-line studies commonly use residual
depth, equivalent width, absorption-weighted velocity, maximum velocity, and
line width to describe outflow profiles and their velocity distributions
\citep[e.g.,][]{martin2005,weiner2009,chisholm2015,chisholm2016}.  For our
stacked spectra we use percentile versions of these quantities, which are less
sensitive to individual noisy pixels than extrema or single-pixel depths.  Let
\(F(v)\) be the continuum-normalized stacked flux as a function of velocity
\(v\) relative to \mgii$\lambda2796$, and define the positive absorption
profile \(D(v)=\max[1-F(v),0]\).  Over \(-700<v<-50\)~\kms, we define the
blue-wing depth as \(1-P_{10}[F(v)]\), where \(P_{10}\) is the 10th percentile
of the normalized flux.  We define width$_{90}$ as the difference between the
95th and 5th percentiles of the velocity distribution weighted by \(D(v)\)
over the same interval.  Thus width$_{90}$ is a robust measure of the velocity
range occupied by the absorbing gas, analogous to the line-width and
\(v_{90}\)-type quantities used in UV absorption-line outflow work
\citep{chisholm2015,chisholm2016}.

For a continuum-normalized resonant absorption profile, the residual flux can
be written schematically as
\begin{equation}
F(v) \simeq 1-C_f(v)\left[1-e^{-\tau(v)}\right]+F_{\rm em}(v),
\end{equation}
where \(C_f(v)\) is the covering fraction of absorbing gas, \(\tau(v)\) is the
optical depth, and \(F_{\rm em}(v)\) represents resonantly scattered emission
that fills the trough.  This standard degeneracy means that the same EW can
arise from a larger column density, larger covering fraction, weaker emission
filling, or absorption extending over more velocities
\citep[e.g.,][]{rubin2011,erb2012,martin2012,martin2013,rubin2014,
chisholm2015}.

The blue-wing absorption depth does not increase monotonically with
$\Sigma_\star$, as shown in the upper panels of Fig.~\ref{fig:mechanism}.  If the EW trend were driven only by increasing covering
fraction at fixed velocity structure and emission filling, the trough would
deepen systematically.  The observed depth behavior disfavors that simplest
interpretation.

In contrast, width$_{90}$ generally increases with $\Sigma_\star$, as shown in the lower panels of Fig.~\ref{fig:mechanism}.  In the
mass+SFR-matched sample, the low-to-high changes are about $38$, $27$, and
$8$~\kms\ in the three redshift bins.  Since the EW increases without a
monotonic increase in trough depth, the additional absorbed area is most
naturally associated with absorption spread over a wider range of velocities,
possibly combined with changes in velocity-dependent covering fraction or
emission filling.

\section{Discussion}

\subsection{Why AGN-clean Samples Are the Appropriate Test}

The AGN-clean selection is designed to isolate star-formation-driven
absorption.  The AGN/composite fraction can increase with $M_\star$,
compactness, and $\Sigma_\star$; retaining such objects could therefore
produce an apparent EW$_{\rm out}$--$\Sigma_\star$ relation unrelated to
stellar feedback.  The trend persists after BPT, \nev, and \heii\ vetoes, and
is already present in the lowest redshift bin where the optical BPT
classification is most complete.

\subsection{Comparison with Previous MgII Work}

Our result complements the DESI \mgii\ census of
\citet{yu2025}.  They showed that cool outflow equivalent width depends
strongly on SFR and that velocity is linked to stellar mass.  We explicitly
control both quantities and find a remaining dependence on
$\Sigma_\star$.  A natural interpretation is that total star formation sets
much of the absorbing-gas strength, stellar mass sets part of the velocity
scale, and compactness controls how efficiently feedback couples to cool gas
at fixed mass and SFR. 

The interpretation is consistent with earlier down-the-barrel absorption work
showing that \mgii\ absorption strength mixes covering fraction, velocity
spread, saturation, and resonant emission filling
\citep[e.g.,][]{martin2005,weiner2009,rubin2014}.  It is also consistent with
the DESI \mgii\ analysis of \citet{yu2025}.  Because \mgii\ is a resonant
transition, an increase in EW$_{\rm out}$ can reflect absorbing gas covering
more velocity space, a larger effective covering fraction, different emission
filling, or some combination of these effects.

\subsection{Physical Picture}

At fixed $M_\star$ and SFR, increasing $\Sigma_\star$ means the same stellar
mass is packed into a smaller projected radius.  In real galaxies this
structural coordinate can trace several coupled physical conditions: a more
compact stellar potential, a different gas column structure, and more
concentrated feedback sites.  The observed response is a larger absorption EW
with a broader blue wing.  High-$\Sigma_\star$ galaxies therefore appear to
place cool gas over a wider range of outflow velocities and/or cover a larger
effective fraction of the stellar continuum.

Dust reddening has little
effect on the locally continuum-normalized absorption profile, though it can
affect the SFR proxy used for matching.  Inclination and morphology may add
scatter to down-the-barrel absorption strengths.  Full radiative-transfer
modeling would be required to convert the empirical depth and width
diagnostics into unique covering fractions, velocity fields, and column
densities.

\subsection{Robustness of the Structural Trend}

The two controlled samples test different aspects of the same trend.  The
$M_\star$-matched sample keeps the observed covariance between surface density
and star formation.  The $M_\star$+SFR-matched sample tests for a residual
structural signal after removing that covariance.  The EW$_{\rm out}$ trend is
present in both.

The bootstrap uncertainties show that the EW$_{\rm out}$ increase is much
more significant than the velocity increase.  In the mass+SFR-matched stacks,
the low-to-high-$\Sigma_\star$ EW differences are positive at
$3.7$--$6.3\sigma$ if the 16th--84th percentile half-widths are treated as
Gaussian errors, whereas the corresponding velocity differences are weak.
The same statement holds in the $M_\star$-matched stacks, where the EW
increase is significant at $5.8$--$7.6\sigma$.  The first two redshift bins,
where the median redshift range across $\Sigma_\star$ tertiles is only
$0.001$--$0.010$, show the strongest EW trends; the broader $0.60<z<1.00$
bin also shows a positive trend.  Thus the result is not driven by a single
redshift interval or by the broadest high-redshift bin.

The redshift cut, $|c\Delta z/(1+z)|<100$~\kms\ with
\(\Delta z=z_{\rm catalog}-z_{\rm DESI}\), limits artificial profile
broadening from catastrophic redshift mistakes before stacking.
Because the normalized spectra are median-stacked with equal galaxy weight,
the measurement is not luminosity weighted.  The galaxy-level bootstrap tests
the sensitivity of each stack to the finite set of contributing galaxies.

\begin{figure}
\centering
\includegraphics[width=0.48\textwidth]{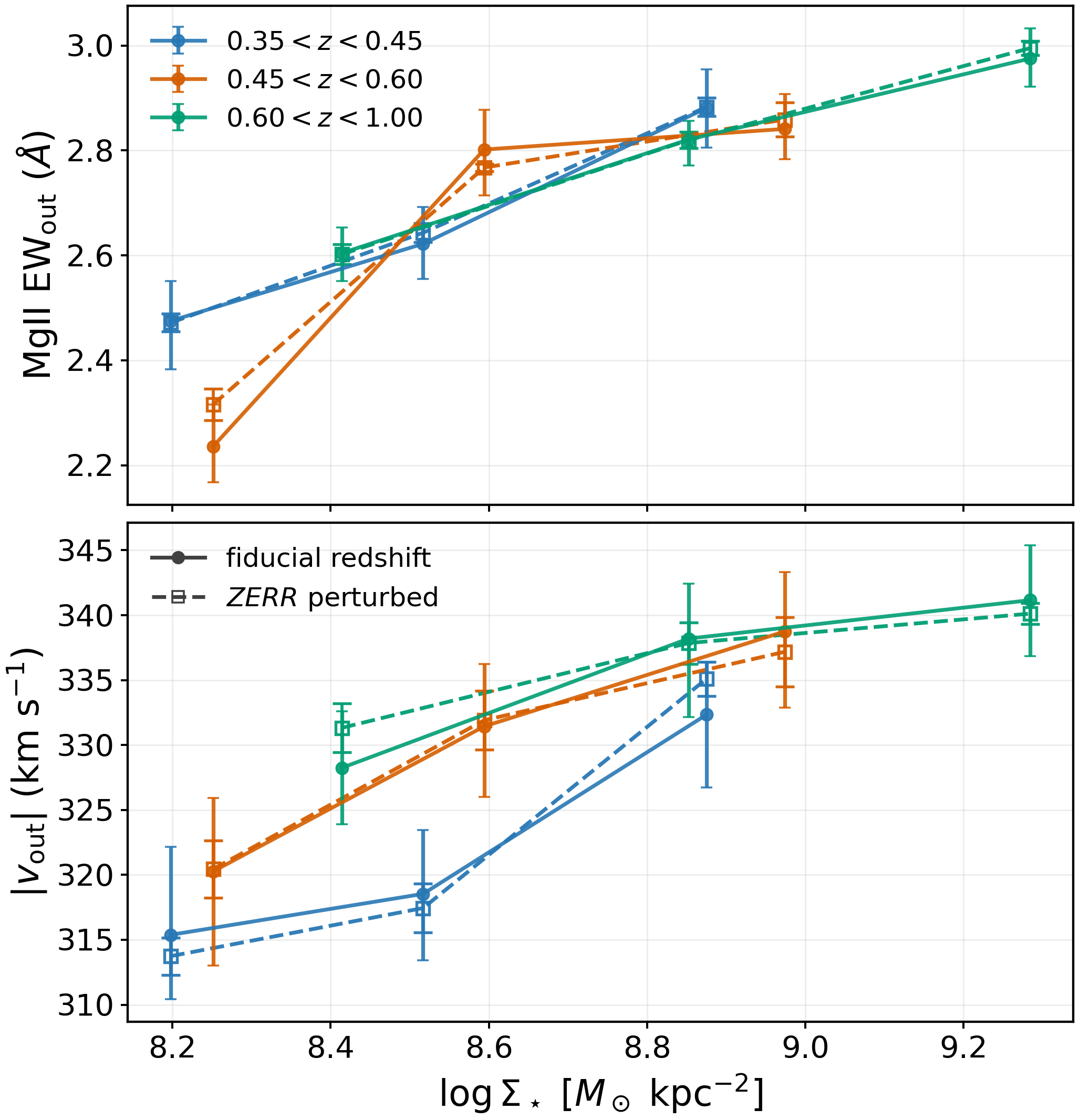}
\caption{Test of statistical DESI redshift errors for the
\(M_\star\)+SFR-matched sample.  Circles show the fiducial stacks using the
official DESI redshifts, and squares show stacks after perturbing each galaxy
by its catalog redshift uncertainty before rest-frame stacking.  The
low-to-high-\(\Sigma_\star\) EW$_{\rm out}$ trend is unchanged within the
redshift-perturbation scatter.}
\label{fig:zerrtest}
\end{figure}

We also tested the effect of the statistical DESI redshift uncertainty on the
most restrictive \(M_\star\)+SFR-matched sample.  For each contributing
galaxy, we perturbed the official DESI redshift by a Gaussian deviate with
standard deviation equal to the catalog redshift uncertainty, shifted the
spectrum to the perturbed rest frame, repeated the stack, and remeasured the
\mgii\ quantities.  Figure~\ref{fig:zerrtest} compares these redshift-perturbed
stacks with the fiducial stacks.  The effect is small.  The perturbed
low-to-high \(\Sigma_\star\) EW$_{\rm out}$ increases are
\(0.41\pm0.01\), \(0.54\pm0.05\), and \(0.39\pm0.03\)~\AA\ in the three
redshift bins, consistent with the fiducial increases of \(0.41\), \(0.61\),
and \(0.37\)~\AA.  The corresponding changes in \(|v_{\rm out}|\) are only a
few \kms.  Thus the statistical redshift uncertainty broadens the profiles
slightly but does not erase the EW$_{\rm out}$--\(\Sigma_\star\) trend.

\begin{figure*}
\centering
\includegraphics[width=0.94\textwidth]{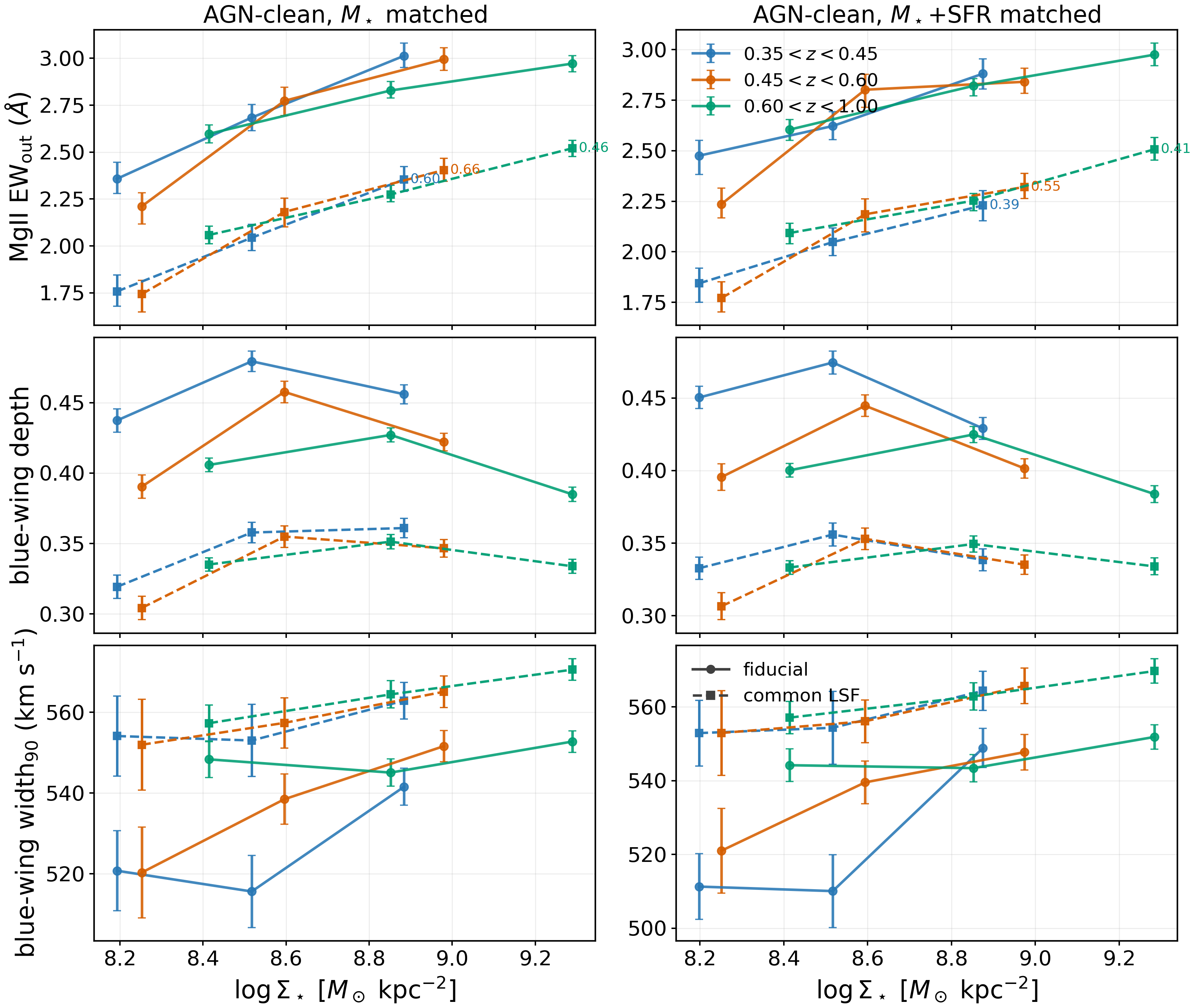}
\caption{Test of the DESI LSF effect on the \mgii\ EW$_{\rm out}$ trend and
on the profile-shape diagnostics used in Figure~\ref{fig:mechanism}.  Circles
show the fiducial stacks used in the main analysis, while squares connected
by dashed lines show stacks after all spectra are convolved to a common
rest-frame LSF.  The common-LSF stacks have smaller absolute EW$_{\rm out}$,
but the low-to-high-$\Sigma_\star$ increase remains in all three redshift
bins and in both controlled samples.  The qualitative profile interpretation
also remains: the blue-wing depth is not a purely monotonic driver, whereas
width$_{90}$ increases with $\Sigma_\star$.}
\label{fig:lsftest}
\end{figure*}

The DESI line-spread function (LSF) is another relevant limitation for stacked
\mgii\ profiles.  The LSF varies with observed wavelength and therefore with
redshift for a fixed rest-frame feature; stacking spectra over a redshift
range also combines galaxies with different instrumental broadening.  The
main quantity in this paper, EW$_{\rm out}$, is a broad-window integral of the
continuum-normalized absorption profile.  A normalized LSF primarily
redistributes absorption area in wavelength, and therefore should not dominate
an EW measurement when the continuum normalization is reliable and the
measurement window is wide enough to include the trough.  The effect is more
important for detailed profile-shape quantities, because LSF smoothing
broadens and shallows absorption features.

For this reason, and to keep the measurement directly comparable to the
empirical DESI \mgii\ stacking analysis of \citet{yu2025}, our fiducial stacks
do not deconvolve individual spectra or convolve all spectra to a common
rest-frame LSF.  \citet{yu2025} likewise treated the DESI LSF as an
interpretive limitation on stacked \mgii\ profile shapes rather than as a
fiducial correction to the stacked spectra.  In our analysis, quantities that
depend on detailed profile shape, especially \(v_{\rm out}\), depth, and
width$_{90}$, should therefore be interpreted as observed stacked-profile
measurements rather than intrinsic gas-velocity widths.

We nevertheless performed an explicit test to evaluate the possible LSF
impact on our structural trends.  For each contributing
spectrum, we estimated the DESI resolution at the observed \mgii\ wavelength,
converted it to the rest frame, convolved the locally normalized rest-frame
spectrum to a common rest-frame Gaussian width of
\(\sigma_{\rm LSF}=2.0\)~\AA, and repeated the stacking and EW measurement.
We then remeasured EW$_{\rm out}$, blue-wing depth, and width$_{90}$ in the
common-LSF stacks.  Figure~\ref{fig:lsftest} compares the fiducial and
common-LSF measurements.  The common-LSF smoothing lowers the absolute
EW$_{\rm out}$ values, as expected for fixed finite measurement windows, but
the structural trend remains.  In the \(M_\star\)-matched stacks, the
low-to-high \(\Sigma_\star\) EW$_{\rm out}$ increases are \(0.60\), \(0.66\),
and \(0.46\)~\AA\ in the three redshift bins.  In the \(M_\star\)+SFR-matched
stacks, the corresponding increases are \(0.39\), \(0.55\), and
\(0.41\)~\AA.  Thus the EW$_{\rm out}$--$\Sigma_\star$ relation is not
produced by redshift-dependent DESI resolution.

The LSF test also leaves the qualitative profile interpretation unchanged.
After common-LSF smoothing, the blue-wing depth still does not provide a
simple monotonic explanation for the EW trend, while width$_{90}$ increases
from the low- to high-\(\Sigma_\star\) tertile in all three redshift bins for
both controlled samples.  The common-LSF width$_{90}$ increases are
\(9\)--\(13\)~\kms\ in the \(M_\star\)-matched stacks and
\(11\)--\(13\)~\kms\ in the \(M_\star\)+SFR-matched stacks.  We therefore
interpret the EW trend as an increase in absorbed velocity extent and/or
velocity-dependent covering and emission filling, rather than as a purely
LSF-driven profile-shape artifact.  The profile widths in
Figure~\ref{fig:mechanism} are still used only to guide the physical
interpretation of the EW trend, not as a precision measurement of intrinsic
velocity dispersion.

\section{Summary}

We have used DESI DR1 stacked spectra to test whether stellar surface density
is an independent empirical coordinate of \mgii\ cool-gas outflows in massive
star-forming galaxies after controlling for stellar mass and SFR.  Our main
conclusions are:
\begin{enumerate}
\item In AGN-clean samples, \mgii\ EW$_{\rm out}$ increases monotonically with
$\Sigma_\star$ in three redshift bins from $0.35<z<1.0$.
\item The trend persists in the fiducial sample matched in both $M_\star$ and
line-SFR, demonstrating that compactness carries information beyond total
stellar mass and our line-SFR proxy.
\item The profile-shape diagnostics suggest that the structural trend likely
reflects changes in the absorbing velocity spread and/or effective covering
fraction of the cool outflowing gas.
\end{enumerate}

DESI therefore reveals stellar surface density as a measurable structural
parameter in cool-gas feedback.  The stacked spectra and tabulated
EW$_{\rm out}$ measurements from this work provide a direct benchmark for
models and future surveys that connect cool-gas covering fraction, absorbing
velocity structure, and galaxy compactness.

\begin{acknowledgments}
This work uses data from DESI.  We thank
Hu Zou for providing the DESI stellar-mass and emission-line
catalog that enables the sample selection used here. YR acknowledges supports from the CAS Pioneer Hundred Talents Program (Category B), the NSFC grants 12522302 and 12273037, and the USTC Research Funds of the Double First-Class Initiative. This work is supported by the China Manned Space Program with grant no. CMS-CSST-2025-A06 and CMS-CSST-2025-A08.
\end{acknowledgments}

\end{document}